\documentclass{llncs}

\usepackage{graphicx}
\usepackage{url}

\newcommand{\code}[1]{{\ttfamily #1}}
\newcommand{\hide}[1]{}

\begin{document}

\title{
Towards an Application of \\
Update Propagation on Logic Programs \\
Representing Java Source Code}
\author{Richard Tantius, Daniel Speicher, Andreas Behrend}

\institute{Computer Science III\\
University of Bonn, Germany}

\institute{University of Bonn, Computer Science III,\\
R\"omerstra\ss{}e 164, 53117 Bonn, Germany\\
\{tantius, dsp, behrend\}@cs.uni-bonn.de}

\date{17 May 2012}
% Just remember to make sure that the TOTAL number of authors
% is the number that will appear on the first page PLUS the
% number that will appear in the \additionalauthors section.

\maketitle
\begin{abstract}
Logic programs are now used as a representation of object-oriented source code
in academic prototypes for about a decade. This representation allows a clear
and concise implementation of analyses of the object-oriented source code. The
full potential of this approach is far from being explored. In this paper, we
report about an application of the well-established theory of update propagation
within logic programs.
Given the representation of the object-oriented code as facts in a logic
program, a change to the code corresponds to an update of these facts. We
demonstrate how update propagation provides a generic way to generate
incremental versions of such analyses.
\end{abstract}

%\terms{Theory}


\section{Introduction}

In this paper, we show how update propagation can be employed for efficiently computing-derived information within the domain of logic fact-bases representing object-oriented source code.

\subsection{Logic Meta Programming}

Logic Meta Programming approaches build a detailed representation of a program (typically) written in another programming language in a logic fact-base \cite{wuyts01}, \cite{kniesel07}, \cite{hajiyev06}.
This representation allows analysing the orginal program by means
of a logic program built on those facts. This approach has
been used to detect code locations that need design improvement
\cite{speicher11}. It provides a suitable basis for different static analyses from the implementation of code quality metrics to
the implementation of type constraints. Recently, we have
been arguing that logic meta programming should be used
to integrate the knowledge about good design structures and
suspicious design structures, creating a database of code quality knowledge, which can be evolved over time \cite{speicher11}.

\subsection{Update Propagation}
Update propagation (UP) is an established database research topic, which has been studied over the last 40 years mainly in the context of integrity checking and materialized
views maintenance, e.g., \cite{BDM88}, \cite{BM04}, \cite{kuc91}, \cite{griefahn97}, \cite{man94}, \cite{oli91}.
Therefore, UP is known from the SQL and Datalog world.
UP makes a contribution to efficiently compute implicit changes of derived relations resulting from explicitly performed updates of extensional facts of a logic fact-base.
SQL view specifications and Datalog rules, as well as Prolog rules are related forms of deductive rules.
This supports the idea to adapt UP to the Prolog world, and also use sets of deltas, together with specialized update statements to incrementally maintain derived predicates (which can be a software analysis, implemented in Prolog).
The original predicates are required only once for materializing their initial answers, the specialized delta versions are used in update statements afterwards for continuously updating the materialized results.
Assuming that a great portion of the materialized content of the original logic rules remains unchanged, the application of such update statements may considerably enhance the efficiency of computing the state of such relations after an update of the fact-base.

\subsection{Refactoring Impact Prediction}

UP enables computing the results of an analysis after an update of the fact-base without actually executing the change.
Such an update of the fact-base can be induced for example by applying a structural improvement like a refactoring \cite{fowler99} to the source code.
This allows to efficiently execute a "what-if"-style analysis.
The approach enables us to evaluate a potential refactoring of Java programs, by computing quality attributes like software metrics before the refactoring and simulate via UP how the metric result changes due to the refactoring.
%In this domain we discuss the refacotrings move field and move attribute  within the scope of this paper.

\subsection{Automation}
\begin{figure}[tp]
\begin{center}
  \includegraphics[trim = 0mm 90mm 30mm 0mm, clip, width=3.1in]{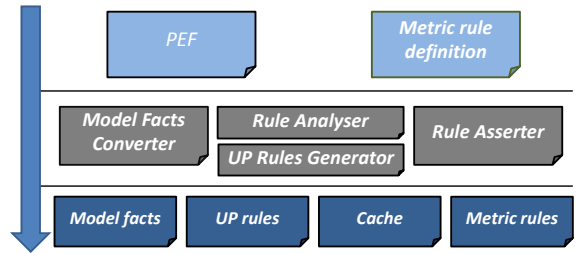}
  \caption{Overview Update Propagation }
  \label{fig:overviewImplementation}
\end{center}
\end{figure}

Figure~\ref{fig:overviewImplementation} introduces our implementation which automates several aspects of the refactoring impact prediction via UP. The picture shows the different components of the system. Those components may consist of several Prolog modules.
The application covers the derivation of a suitable abstract model, on which we build the software analysis we intend to employ.
We use the Logic Meta Programming approach JTransformer presented in \cite{kniesel07}, to derive such a model. The implementation also handles the refactoring simulation on the level of that abstract model, metric computation and UP rule generation. We have used SWI-Prolog (in the version 6.0.2\footnote{The Project homepage of SWI-Prolog:
\url{http://www.swi-prolog.org/} (accessed
17.08.12).}) for the implementation. Beside the metric definitions that had to be implemented in a strictly declarative syntax, so that we can apply UP, we used the full feature set of the SWI-Prolog environment for the other parts of the implementation.

% Fig. 2 illustrates this idea.

%Limited (EXPLAIN!) validity of metrics, Interactive Refactoring plan creation

%Combination of Metrics, Search based refactoring

\section{Model and Analysis}
\label{section:approach}
\hide{
Our approach consists of several steps. First, we need to choose an appropriate
model for our problem domain in order to specify the software metrics, to which
we will apply the precomputation approach.
As mentioned before we will implement the model and the metric computation in
Prolog.
To describe the refactoring impact on the model, for each refactoring it then is
necessary to formulate the following:

\begin{enumerate}
\item The refactoring signature determines the input parameters of a
refactoring. For example effected elements like classes, method or fields.
\item The precondition specifies when a refactoring may be applied to the source
code.  % Not the model!!
\item The postcondition describes the assertions that can be made ​​with respect
to our model, after the refactoring has been applied.
 \end{enumerate}

From the postcondition we then derive the changes on the model in terms of delta
facts, representing added and deleted model facts.
To make the metrics capable of precomputation we apply update propagation on
their computation rules. Actually the application of update propagation only has
to be done once for each metric.
}

Software analyses like software metrics are a frequently-studied approach to detect lack of quality and are also capable of making improvements of quality measurable.
Logic Meta Programming Approaches provide the capability to represent software systems as logic programs.
A refactoring in this context, therefore, can be understood as a transformation imposing changes on an extensional fact-base.
We discuss the structural cohesion metric Lack of Cohesion in Methods $LCOM1$ \cite{chidamber94}, as an example of such a software analysis.
Cohesion can be defined as the degree of how closely module components are related to each other.
A unified framework for structural metrics was presented by Briand in \cite{briand99}, who created a common model for existing metrics. The model unified the syntactical representation and operational semantic of those metrics.
In order to provide the information about the source code the metric relies on, we also present a simplified abstract model as basis for $LCOM1$. This meta model will be directly derived from the Logic Meta Programming fact-base.

\subsection{Abstract Cohesion Model}

The Logic Meta Programming approach \cite{kniesel07}, which we use to derive our abstract cohesion model, is based on Prolog. For this reason, we represent the relevant information for the $LCOM1$ metric as Prolog facts. We will also present the $LCOM1$ metric itself as a logic program in the following.
To sufficiently describe the information relevant for cohesion metric, we need to take the following information into account: \textit{Which class contains a certain method or field? Which methods are
called and which fields are accessed by a method?} The presented model is based on the cohesion model as presented in Briand \cite{briand99} and was adapted to Prolog.
We consider the following Prolog predicates:

\begin{tabular}{ll}
\verb|  c(M).|&\verb|  |class\\
\verb|  cm(C, M).|&\verb|  |class contains method\\
\verb|  cf(C, F).|&\verb|  |class contains field\\
\verb|  mf(M, F).|&\verb|  |method accesses field\\
\verb|  mm(M, N).|&\verb|  |method invokes method\\
\end{tabular}

The related Prolog facts are ground versions of the predicates from above and the variables \verb|C,M,F,...| are bound to unique identifiers for the corresponding elements. In the next subsection, we demonstrate how we extract this model from the JTransformer fact-base.
In the following section, we build the $LCOM1$ metric on top of those facts as a logic program.
\hide{
\begin{figure}[h]
\begin{center}
\includegraphics[page=2,width=0.48\textwidth]{../img/Overview}
  \caption{Abstracting from a concrete Refactoring to explore the impact of the
  refactoring on structural metrics.}
  \label{fig:abstraction}
\end{center}
\end{figure}
}

\subsection{Model Fact Derivation}

We derive the abstract cohesion model directly from the fact-base created by the JTransformer Logic Meta Programming approach. JTransformer builds a so called \textit{Abstract Syntax Tree} (AST) that already is a full model representation in Prolog of the Java language. The JTransformer facts are called \textit{Program Element Facts} \textbf{PEF}, they are the starting point for creating the cohesion model.

Figure~\ref{fig:overviewImplementation} gives an overview of the various  components of our Implementation. The architecture consists of three layers. In the first layer, we have the metric rules and the facts created by JTransformer. The layer in the middle is the meta programming layer, in which we process the rules and facts. Here we compile the UP rules. In the layer at the bottom, we have the subsystems which actually contain the executable code. Each subsystem may consist of several Prolog modules.

Based on generator predicates, the \textbf{Model Facts Converter} component creates cohesion model facts (from the JTransformer facts) and asserts them to a prolog module (\verb|cohesion_model|). Additionally, we perform checks, if an element should be included at all. In the case of classes, we do not consider the three following class types. \textit{Interface classes} do not provide method calls and attribute references. Cohesion cannot be examined here. \textit{Classes from external dependencies} are supposed to be examined elsewhere. \textit{Anonymous classes} are not supposed to be analysed standalone.

The implementation of the class generator predicates is as follows:
\begin{verbatim}
  generate(FactsModule, c, [ClassId]) :-
    % Here we only use the class id
    classT(ClassId, _, _, _),
    source_class(ClassId).
\end{verbatim}
% This was a quickfix, not relevant for the paper.
%,
%    % Redundancy check
%    not(FactsModule:c(ClassId)).
%
For the free variable \verb|FactsModule|, we use the module \verb|cohesion_model| mentioned above as a default value. The implementation of the class analysis considers different JTransfomer facts to determine the class type:
\begin{verbatim}
  source_class(ClassId) :-
    % JTransformer facts
    not(externT(ClassId)),
    not(interfaceT(ClassId)),
    % See below
    not(anonymous_class__(ClassId)).
\end{verbatim}
\begin{verbatim}
  anonymous_class__(Class):-
    classT(Class, _, ClassName, _),
    string_concat('ANONYMOUS$', _, ClassName).
\end{verbatim}
On top of the provided model facts, we create our software analyses, for example the cohesion metric presented before. The deductive \textbf{Metric rule definition} is the starting point (also shown in Figure~\ref{fig:overviewImplementation}). We define the deductive part of the metric in a Prolog module:

\verb| :- module(| $metricName$\verb|_deductive_ruleset, []).|\\
\verb|       |$rule_{1}$\verb|(A, B, |$...$\verb|) :- |$...$\\
\verb| |\\
For example:\\
\verb|    :- module(lcom1_deductive_ruleset, []).|\\
\verb|       ...|\\

\subsection{Declarative Metric Implementation}
The $LCOM1$ metric definition we present in the following is based on the definition given by Briand in \cite{briand99}.

\subsubsection{Query and Mapping}

The computation of structural metrics can be divided into two steps. First, a query step collects the elements or relations that are relevant for the metric.
Second, a mapping maps the result of the query to a number. Separating these
steps has the benefit that we can discuss both steps on their own. A query
result may be evaluated with different mappings. A mapping may be applied to the
result of different queries. The
separation of query and mapping has the advantage, that we can apply the update
propagation approach to the deductive rules defining the query.

\paragraph*{Query} $LCOM1$ counts the number of
method pairs within a class that do not access even one common field. We split this definition into two predicates. Each
predicate will be first described in natural language, then as a logic program.

\vspace{0.5em} \noindent {\em $M$ and $N$ in $C$ are connected, if $M$ accesses
a field $F$, that belongs to the class $C$, and $N$ accesses that field $F$
as well.}
\begin{verbatim}
  cp(C, M, N) :-
      mf(M,F), cf(C, F), mf(N,F).
\end{verbatim}

\vspace{0.5em} \noindent {\em $M$ and $N$ in $C$ are a pair of methods lacking
cohesion, if $M$ is a method in $C$, $N$ is as well a method in $C$ and $M$ and
$N$ are not a connected pair in $C$: }

\begin{verbatim}
  lp(C, M, N) :-
      cm(C, M), cm(C, N), not(cp(C, M, N)).
\end{verbatim}

\paragraph*{Mapping}

To complete the $LCOM1$ computation rule, we need to perform some
additional steps after the deductive part.

\begin{verbatim}
  lcom1(C, R):-
    findall([M,N], (cp(C, M, N), not(M=N)), E),
    length(E, T),
    R is T/2.
\end{verbatim}

\section{Refactoring as Cohesion Model Update}
\label{section:refAsCohModelUp}

We model a refactoring as an update on our cohesion model.
Performing the refactoring only on the abstract level of the cohesion model, helps to concentrate the computation only on aspects relevant for the metric computation. Because we use update propagation in the following, we do not directly perform model updates, rather every update will generate a so called delta fact, which depicts the actual change and will be discussed in detail in Section~\ref{section:updatePropagation}.
In the following, we briefly discuss the refactorings we use and describe their effects on the level of the model.
Both refactorings assume in our setting, that the moved element will be extracted into a new class, creating the following delta fact: \verb|add_c(#newClassId)|

Since we exclude constructor methods from our model and do not consider modificators as public, private and protected, the following refactorings require no special preconditions to be applied.

\subsection{Move Method, Move Field}
The move method refactoring moves a method from one class to another. Though in a real world refactoring, we would need to adjust the code in several ways, so that it remains functioning and compileable we add two simple delta facts to our model:
\begin{verbatim}
  add_cm(C, M)
  del_cm(C, M)
\end{verbatim}
Similar to the move method refactoring the move field refactoring moves a field from one class to another, the resulting delta facts are as follows:
\begin{verbatim}
  add_cf(C, F)
  del_cf(C, F)
\end{verbatim}

\section{Update Propagation}
\label{section:updatePropagation}

In this section, we show how to apply update propagation in Prolog.
Because of the different evaluation mechanisms for SQL views (set-oriented, bottom-up) and Prolog rules (instance-oriented, top-down), however, the transformation techniques from UP could not
be applied directly. Instead, the specific properties of Prolog rules
have to be taken into account in order to achieve a complete and sound
update propagation based on delta predicates.

\subsection{Rule Transformation in Prolog}
\label{section:ruleTransformationInProlog}

%\begin{figure}[tp]
%\begin{center}
%  \includegraphics[trim = 17mm 100mm 23mm 83mm, angle=90, clip, %width=4.8in]{../img/tccRulesClean.pdf}
%  %\caption{The derived update propagation and indirect and direct %transition rules for the $TCC$ definition.}
%  \label{fig:ruleTransformationInProlog}
%\end{center}
%\end{figure}
%\section{Applying Update Propagation}

\begin{figure}[t]
\begin{center}
{
\fontsize{6}{4}
\begin{verbatim}
% Derived from: lp(C, M, N) :- cm(C, M), cm(C, N), not(cp(C, M, N)).

add_lp(C, M, N) :- add_cm(C, M),                  nwd_cm(C, N), not(nwi_cp(C, M, N)), not(    lp(C, M, N)).
add_lp(C, M, N) :- add_cm(C, N),    nwd_cm(C, M),               not(nwi_cp(C, M, N)), not(    lp(C, M, N)).
add_lp(C, M, N) :- del_cp(C, M, N), nwd_cm(C, M), nwd_cm(C, N),                       not(    lp(C, M, N)).

del_lp(C, M, N) :- del_cm(C, M),                      cm(C, N), not(    cp(C, M, N)), not(nwi_lp(C, M, N)).
del_lp(C, M, N) :- del_cm(C, N),        cm(C, M),               not(    cp(C, M, N)), not(nwi_lp(C, M, N)).
del_lp(C, M, N) :- add_cp(C, M, N),     cm(C, M),     cm(C, N),                       not(nwi_lp(C, M, N)).

nwi_lp(C, M, N) :-                  nwd_cm(C, M), nwd_cm(C, N), not(nwi_cp(C, M, N)).

nwd_lp(C, M, N) :-     lp(C, M, N), not(del_lp(C, M, N)).
nwd_lp(C, M, N) :- add_lp(C, M, N).



% Derived from: cp(C, M, N) :- mf(M, F), cf(C, F), mf(N, F).

add_cp(C, M, N) :- add_mf(M, F),               nwd_cf(C, F), nwd_mf(N, F), not(    cp(C, M, N)).
add_cp(C, M, N) :- add_cf(C, F), nwd_mf(M, F),               nwd_mf(N, F), not(    cp(C, M, N)).
add_cp(C, M, N) :- add_mf(N, F), nwd_mf(M, F), nwd_cf(C, F),               not(    cp(C, M, N)).

del_cp(C, M, N) :- del_mf(M, F),                   cf(C, F),     mf(N, F), not(nwi_cp(C, M, N)).
del_cp(C, M, N) :- del_cf(C, F),     mf(M, F),                   mf(N, F), not(nwi_cp(C, M, N)).
del_cp(C, M, N) :- del_mf(N, F),     mf(M, F),     cf(C, F),               not(nwi_cp(C, M, N)).

nwi_cp(C, M, N) :-               nwd_mf(M, F), nwd_cf(C, F), nwd_mf(N, F).

nwd_cp(C, M, N) :-     cp(C, M, N), not(del_cp(C, M, N)).
nwd_cp(C, M, N) :- add_cp(C, M, N).



% Direct transition rules for the model facts mf/2 and cf/2

nwd_mf(C, M) :- mf(C, M), not(del_mf(C, M)).
nwd_mf(C, M) :- add_mf(C, M).

nwd_cf(C, F) :- cf(C, F), not(del_cf(C, F)).
nwd_cf(C, F) :- add_cf(C, F).
\end{verbatim}
}
\end{center}
\caption{The derived update propagation and indirect and direct transition rules for the $LCOM1$ definition.}
\label{fig:ruleTransformationInProlog}
\end{figure}

The task of UP is to systematically compute the set of all
induced changes, starting from the physical changes of base data.
Technically, this is a set of delta facts for any affected predicate
which may be stored in corresponding delta relations. For each
predicate symbol \code{p}, we will use a pair of delta predicates
\code{<add\_p, del\_p>}
representing the insertions and deletions induced on \code{p} by an
update. The initial set of delta facts represents the so-called
\emph{UP seeds}.

In the following, we briefly review a transformation-based approach
to UP where the Prolog rules and the UP seeds are employed to
derive \emph{propagation rules} for computing delta relations.
A propagation rule refers to at least one delta predicate in its body
in order to provide a focus on the underlying changes when
computing induced updates. For showing the effectiveness of an
induced update, however, references to the state of a predicate
before and after the base update has been performed are necessary.

For each predicate ${\tt p}$ we use ${\tt old\_p}$ to refer to its old state before the changes given in the delta sets have been applied (technically the rule behind ${\tt old\_p}$ is the unmodified version of ${\tt p}$). We use ${\tt new\_p}$ to refer to the new state of ${\tt p}$.
These state relations are never completely computed, but are queried
with bindings from the delta sets in the propagation rule body, and
thus act as a test of effectiveness.
An induced insertion or induced deletion can be simply represented by the difference between the two consecutive database states.
We consider the following Prolog rule:
\begin{verbatim}
  p(X) :- q(Y),r(Z),not s(C).
\end{verbatim}
The difference rules may look as follows:
%\begin{scriptsize}
\begin{verbatim}
  add_p(X) :- add_q(Y), new_r(Z), not(new_m(C)), not(old_p(X)).
  add_p(X) :- new_q(Y), add_r(Z), not(new_m(C)), not(old_p(X)).
  add_p(X) :- new_q(Y), new_r(Z),     del_m(C),  not(old_p(X)).
\end{verbatim}
%\end{scriptsize}
The propagation rules basically perform a comparison of the \verb|old| and
\verb|new| versions of the predicates. While providing a focus on insertions into
 q and r, all necessary combinations of delta and state predicates are considered.
Because of the negative referenced predicate s, an additional rule has to be considered,
which covers new derivations for p due to a deletion from s.

All propagation rules also contain the additional effectiveness test \verb|not old_p(X)|, to check for the effectiveness of the induced insertions in case of alternative derivations of the same fact \verb|p| in the old state.
As an optimization we can drop the test, in case there are no alternative derivations, or if the set of, insertions can be overestimated. To avoid the full determination of state predicates, we should move the delta predicates as far left as possible in the rule body. This way the bindings provided by the delta facts can be used for restricting the evaluation of state predicates. This leads to the following rules:
\begin{verbatim}
  add_p(X) :- add_q(Y), new_r(Z), not(new_m(C)).
  add_p(X) :- add_r(Z), new_q(Y), not(new_m(C)).
  add_p(X) :- del_m(C), new_q(Y),     new_r(Z).
\end{verbatim}

For simulating the new predicate state from a given
update and the old state, so called \emph{transition rules}~\cite{oli91}
can be used. The transition rules of a derived predicate infer its new state from the new states of the underlying predicates. Thus, for a rule $A
:- L_1,\ldots, L_n$ a transition rule of the form
$new\_A :- new\_L_1,\ldots,new\_L_n$ is considered. In contrast,
for every extensional predicate $A$ so-called incremental transition rules are used:

\begin{verbatim}
  new_A :- old_A, not(del_A).
  new_A :- add_A.
\end{verbatim}
which explicitly refer to the computed changes to \code{A}.

For a rule $A:- L_1,\ldots, L_n$ we may also use the direct transition rules if there are no mutual dependencies between the predicate and the predicates in the body. In any case, the indirect transition rules of the form\\ \verb|new|$\_A$\verb| :- new|$\_L_1$\verb|,|$\ldots,$\verb|new|$\_L_n$\\ are used in the effectiveness test of negative propagation rules.

As an example, consider the following Prolog program
\begin{verbatim}
  p(X) :- q(X,Y),r(Y),not(s(Y)).
          q(1,2). r(3). s(4).
          q(2,3). r(4). s(5).
          q(3,4). r(5). s(6).
\end{verbatim}
and the insertion \verb|r(2)| into relation r. The following
propagation and transition rules
\begin{verbatim}
  p(X) :- add_r(Y),new_q(X,Y),not(new_s(Y)).
  new_q(X,Y) :- q(X,Y), not(del_q(X,Y)).
  new_q(X,Y) :- add_q(X,Y).
  new_s(X,Y) :- s(X,Y), not(del_s(X,Y)).
  new_s(X,Y) :- add_s(X,Y).
\end{verbatim}
were derived using the scheme described above. These rules allow for
efficiently computing the induced insertion p(1) (represented
by the fact $add\_p(1)$) by avoiding any redundant recomputations.

\hide{
\subsection{Performance Aspects}
Update propagation does not always lead to an improved evaluation.
In general, the efficiency of UP strongly depends on the number
of avoided recomputations with respect to derived facts which are
unaffected by the underlying refactoring steps. Thus, if the
number of derived facts that remain to be derivable
is considerably higher than the number of induced changes, then
UP can indeed lead to more efficient consequence analysis. On the
other hand, the  maintenance of delta sets as well as the application
of complex update statements introduces some overhead to the system.
If the number of avoided recomputations is relatively low in
comparison to the number of propagated updates then UP may even
perform worse than a complete recompuation of all derived facts due
to this overhead. Therefore, various methods for reducing the
overhead induced by UP have been proposed in the literature
(such as Magic Updates~\cite{behrend04}) which allow for
broadening the application spectrum of UP even more.
}
\section{Propagation of Cohesion Model Updates}

In this section, we apply UP as presented in Section~\ref{section:updatePropagation} to our cohesion model and the metric rules from Section~\ref{section:approach}.

Before we employ UP to simulate the model state and metric results after the refactoring, we first need to generate delta facts as described in Section~\ref{section:refAsCohModelUp}. At definition time of the metric rules, we may also derive the propagation and transition rules for UP. Figure~\ref{fig:ruleTransformationInProlog} shows the result of the rule derivation process. We can see that for each body literal of a rule that appears in the metric definition, we create a positive and a negative propagation rule, so that for the cohesive pair rule \verb|cp| we obtain six propagation rules in total.
We replace negated delta literals simply by their opposite versions, for example: \verb| not(del_cp) | $ \Leftrightarrow $ \verb| add_cp|. For both rules \verb|lp| and \verb|cp|, we also see two derived transition rules \verb| nwi_lp | and \verb| nwi_cp|, simulating the versions of those predicates in the new state, which operate on the propagation rules.

\subsection{Rule Generation}

The UP rule generator component consists of two subcomponents, which can be seen in the middle of Figure~\ref{fig:overviewImplementation}. The \textbf{Rule Analyser} component collects meta information about the metric rules from the deductive rule modules and the \textbf{UP Rules Generator}, which creates the UP rules, based on the collected meta information. SWI-Prolog provides various meta predicates to examine loaded programs.
It is important that UP ensures that the augmented rule set (which includes UP rules as shown in Figure~\ref{fig:ruleTransformationInProlog} and the original rules like those for $LCOM1$) generated for a set of logic rules which is guaranteed to terminate, still keeps this property. This was shown for the language set of Datalog \cite{behrend04}, \cite{griefahn97}, which only allows straight forward declarative rules, in comparison to Prolog. We also need to assure this in Prolog, it would be unfavourable if the UP rules got stuck in infinite loops. Prolog allows a broad variety of syntax constructs.
For the metric definition, therefore, we only allow the model predicates and predicates defined in the template module itself and those from the cohesion model. We also allow a narrow list of built-in predicates, namely \verb|=/2|, \verb|member/2| and the negation predicate \verb|not/1|. We also do not allow any complex terms in the head of the rules, like: \verb|  cp(a(A),B,[H|$|$\verb|T]) :- ...|. Though this is a sharp restriction of Prolog, we were able to describe several structural cohesion metrics, as long as they contained a deductive part.

\paragraph*{Rule Analyser} First, the \textbf{Rule Analyser} determines all predicates defined in the template module which contains the $LCOM1$ metric rules we presented before. For $LCOM1$ those are \verb|cp| and \verb|lp|. The analyser collects various meta information, which we need to build the UP rules.
An overview of the collected meta information:\\
\\  \verb| head_of_rule(|$headId$\verb|, |$groundHead$\verb|, [|$name$\verb|, |$arity$\verb|])|\\
\verb| body_predicate_of_rule(|$headId$\verb|, |$bodyId$\verb|,|\\\verb|                  |$positonInBody$\verb|, |$groundPredicate$\verb|)|\\
\verb| rule_variables(|$headId$\verb|, |$bodyId$\verb|, |$headBodyPrefix$\verb|, |\\\verb|                  |$positonInBody$\verb|, |$groundPredicate$\verb|)|\\
\verb| |\\
\verb|  predicate_dependencies_transitive_closure(|\\\verb|                  [|$name$\verb|, |$arity$\verb|], |$dependencies$\verb|)|\\
\verb| |\\
We need to determine the free variables in the head and the body of the rules. We also need to check, if the rules contain self references. This is relevant to determine the positioning of the delta terms (\verb|del_| and \verb|add_| in the body of propagation rules).
Second, in order to create the indirect transition rules \verb|nwi_P| of a predicate \verb|P|, we need to analyse, if there are transitive mutual dependencies between \verb|P| and its body predicates \verb|L|$_i$ (where 0 $\le$ i < \#number of body predicates of \verb|P| ). The result will determine if we rather use \verb|nwi_L|$_i$ or \verb|nwd_L|$_i$ in \verb|nwi_P|, this is important to ensure that the rules still terminate.
We therefore create a predicate dependency graph.

\paragraph*{UP Rules Generator} Based on the information collected by the rule analyser, the \textbf{UP Rules Generator} creates the UP rules.
As mentioned before, we create negative and positive Propagation Rules (for the metric), the direct Transition Rules (metric and model facts) and indirect Transition Rules (metric only).
The rules created are asserted to a module. After completing the generation process, we also compile all rules to static predicates by using the Prolog \verb|compile_predicates| predicate.
The rule creation is based on string concatenation, therefore we convert all predicates to atoms, and prepend the necessary augmentations to each predicate.

\section{Conclusions and Future Work}

Earlier research had shown that the representation of object-oriented source
code as a fact-base in a logic program allows for clear and concise
implementation of static analyses of the object-oriented code. We explored the
potential of this approach further by applying the well-established theory of
update propagation to it. Update propagation gives us a generic way to transform
any analysis represented as a (sufficiently well-formed) logic program into an
incremental version. Given an actual or hypothetical small change to the fact-base, this incremental version of the analysis provides an efficient way to
caculate the new result of the analysis after the (hypothetical) change.

We implemented a transformation, generating the propagation and transition rules
based on the original rules. Besides the applicability of update propagation to
our setting, we already conducted several experiments to explore the performance benefits.
In our experiments, update propagation was significantly faster than actually transforming our model and re-evaluating the original definition. As a future work, we intend a detailed study of the performance benefits.

% Future work

We focused on the precalculation of the refactoring impact on metrics as this is
in line with our research interest. Nevertheless, there is no reason to limit
update propagation for logic meta programming to metrics or not even to
refactoring. In the context of refactoring, update propagation could for example
be used to verify that certain constraints will still hold true after the
refactoring. This would be in line with the original motivation of update
propagation.

A precise definition of the pairs of abstract refactorings on the model and the corresponding refactorings on the source code, could ease and clarify the nature of the induced updates model updates.
More sophisticated refactorings also need complex preconditions, before they may be applied legally. We should be able to check those conditions on the level of sour model.

\hide{

\acks

%[Only if we like the paper, we may allow %ourselves to express our thankfulness]
\begin{itemize}
  \item Cassell for remarks on code
  \item Jan Nonnen for more test cases
  \item Günther, Tobias, JTrf Team
\end{itemize}

}

\bibliographystyle{abbrv}

% The bibliography should be embedded for final submission.

\end{document}